\documentclass[14pt]{iopart}

\usepackage[dvips]{epsfig}
\usepackage{iopams}
\usepackage{setstack}
\newenvironment{align}{\begin{equation}}{\end{equation}}
\newcommand{\mytext}[1]{\rm{#1}}
\newcommand{\mytextrm}[1]{\rm{#1}}
\newcommand{\notag}{\nonumber}

\newcommand{\erf}{\mathop{\mytextrm{erf}}}

\newcommand{\Int}{\mathop{\mathrm{Int}}}

\begin{document}
\title{A frequency measure robust to linear filtering}
\author{A G Rossberg}
\address{Zentrum f{\"u}r Datenanalyse und Modellbindung,
  Universit{\"a}t Freiburg, Eckerstr.~1, 79104 Freiburg, Germany}
\ead{axel@rossberg.net}
\begin{abstract} 
  A definition of frequency (cycles per unit-time) based on an
  approximate reconstruction of the phase-space trajectory of an
  oscillator from a signal is introduced.  It is shown to be invariant
  under linear filtering, and therefore inaccessible by spectral
  methods.  The effect of filtering on frequency in cases where this
  definition does not perfectly apply is quantified.
\end{abstract}

\submitto{\NL, 22 January 2003}


\section{Introduction}
\label{sec:introduction}

In the experimental exploration of complex systems, such es those
encountered in life-science, geology, or astronomy, it is not unusual
that the experimenter discovers oscillations of unknown origin in a
measured time-series.  The experimenter would then usually try to
characterize these oscillations in a form that admits an
identification of their source -- the oscillator.  The conventional
first choice is a characterization of the oscillations by their
``frequency''.  For an ideal, periodically oscillating signal $x(t)$,
the smallest number $T>0$ such that $x(t)=x(t+T)$ for all $t$ is the
\emph{period} of the signal, and its (angular) \emph{frequency} is
defined by $\omega=2\pi/T$.  There is a good reason for choosing this
particular characterization.  All other properties of an ideal,
periodic signal, i.e.\ its waveform and amplitude, are subject to
distortions along the signal pathway from the oscillator to the
detector.  In fact, linear filtering along the signal pathway would
generally be sufficient to modify the waveform and the amplitude in an
arbitrary way.  And the precise properties of the signal pathway are
unknown in the setting considered here.  The frequency information is
the only sure fact.  For ideal, periodic oscillations, these
observations are too obvious to deserve much discussion.  But for
non-ideal oscillations, as they are frequently encountered in complex
systems, the situation is less clear.

A large variety of methods is being used to determine a ``frequency''
for non-ideal oscillations, and not all of them are equally robust to
filtering and other distortions.  Two major kinds of methods can be
distinguished: Firstly, there are \textit{period-counting} methods,
where, from the number of oscillation periods $n(\Delta t)$ in a time
interval $[t_0,t_0+\Delta t]$, the frequency is determined as
\begin{align}
  \label{f_period}
  \omega_{\mytext{count}}=\lim_{\Delta t\to\infty}\frac{2\pi n(\Delta
    t)}{\Delta t}.
\end{align}
(Finite sample-size effects are not discussed here.)  The methods
differ in the criteria used for counting individual periods (e.g., local
maxima, zero-crossings).  Secondly, there are \emph{spectral}
methods, where the frequency $\omega_{\mytext{spec}}$ characterizes
the position of a peak in an estimate of the power spectral density
$S_x(\omega)$ \cite{brockwell91:_time_series,priestley81:_spect} of
the signal $x(t)$ (e.g., \cite{mendez98:_differ_frequen_kiloh_qpos,
  godano99:_sourc_strom_vulcan_islan_isole_eolie_italy,
  korhonen02:_estim_cardiovasc,timmer96:_quant_tremor,
  slavic02:_measur_turbo_wheel}).  Often, the frequency with maximum
power $\omega_{\mytext{peak}}$ is used.  The term \emph{spectral
  methods} shall here also include methods based on estimates of the
autocorrelation function of the signal, since this is related to the
spectral density by a simple Fourier transformation.

For weak distortions and not too irregular oscillations,
period-counting methods can be just as unequivocal as frequency
measurements for ideal oscillations.  This is why they are routinely
used in high-precision frequency (or time) measurements.  They are
also naturally associated with mode-locking phenomena
\cite{pikovsky01:_synch}.  But for stronger distortions and more
irregular oscillations, this robustness is reverted to its contrary.
Unequivocally identifying individual periods of oscillation then
becomes difficult.  In these situations spectral methods are generally
preferred.  However, it is obvious that spectral methods are not
robust to filtering along the signal pathway either.  By linear
filtering, the power spectrum can be modified nearly arbitrarily.

How much can the concept of period-counting frequency measurement be
extended to distorted time series?  A partial answer is given in this
work.  In section~\ref{sec:definition} a generalized period-counting
frequency measure, the \emph{topological frequency}, is defined.  It is
based on the approximate reconstruction of the phase-space trajectory
of the oscillator.  In section~\ref{sec:invariance} it is shown that
the topological frequency is robust with respect to nearly arbitrary
linear filtering.  This has three important implications: (i) At least
as long as the signal pathway acts as a linear filter, the topological
frequency is a characteristic of the (typically nonlinear) oscillator
alone. (ii) Filtering of the signal, in order to remove noise and other
undesirable components, does no harm to the result for the frequency.  In
view of (i) and (ii), the topological frequency can be considered to
be robust with respect to both kinds of distortions, filtering and
noise.   Finally,
(iii) the results of frequency measurements using spectral methods can
deviate arbitrarily from the topological frequency.  This point is
made rigorous in section~\ref{sec:arbitrarity}.

Not for all oscillatory time series can the topological frequency be
defined.  In particular, linear time series driven by Gaussian noise
are excluded.  Weakly nonlinear models for noisy time series can
interpolate between linear Gaussian oscillations and ideal
periodicity.  For these models, the influence of filtering on a weaker
period-counting frequency measure, the average or \emph{phase
  frequency} (see below), is investigated in
section~\ref{sec:imperfect}.  The susceptibility of the phase
frequency to filtering is found to decay rapidly with the degree of
nonlinearity.  Section~\ref{sec:conclusion} contains some concluding
remarks.

\section{Topological Frequency}
\label{sec:ftop}

\subsection{Definitions}
\label{sec:definition}

The theory becomes more transparent in a discrete-time representation.
Let $\{x_t\}$ be an infinite, real-valued time series sampled at
equally spaced times starting at $t=0$.  Measure time in units of the
sampling interval.  Define the spectral density $S_x(\omega)$ of
$\{x_t\}$ as
\begin{align}
  \label{define_power}
  S_x(\omega)=\frac{1}{2 \pi} \sum_{\tau=-\infty}^\infty \left< x_t \,
    x_{t+|\tau|} \right>_t \cos({\omega \tau}),
\end{align}
where $\left<\cdot\right>_t$ denotes temporal averaging ($t \ge 0$).

Let the \emph{trajectory} $p(t)$ of a time series $\{x_t\}$ in
$N$-dimensional delay space be defined by
$p(t)=(x_t,x_{t-1},\ldots,x_{t-N+1})$ for integer $t$ and by linear
interpolation\footnote{In practice, higher order interpolation might
  sometimes be useful.} for non-integer $t$.  Frequency will here be defined
with respect to a Poincar{\'e} section or \emph{counter}, which is
an $(N-1)$-dimensional, oriented manifold $M$ with boundary $\partial
M$ and interior $\Int M=M \backslash \partial M$, embedded in the
$N$-dimensional delay space.

Let $n(t_1)$ be the \emph{oriented number} of transitions of the trajectory
$p(t)$ through $\Int M$ in the time interval $0<t<t_1$.  That is, a
transition through $\Int M$ in positive (negative) direction increments
(decrements) $n(t_1)$ by one.  For example, a positive-slope
zero-crossing counter in $2$-dimensional delay space would be given
by\footnote{Precisely, the atlas containing the single map $m:r\in
  \mathbf{R}^{\ge0} \to (r,-r)\in\mathbb{R}^2$ and an orientation
  defined on it.}
\begin{align}
  \label{zero_crossing}
  M=\{(v_1,v_2)\in \mathbb{R}^2 | v_1+v_2=0,\, v_1\ge v_2\}.
\end{align}

Define the \emph{topological frequency} $\omega_{M,x}$ of $\{x_t\}$
with respect to a counter $M$, as
\begin{align}
  \label{def_top_freq}
  \omega_{M,x}:=\lim_{t\to \infty} \frac{2\pi|n(t)|}{t},
\end{align}
provided the limit exists \emph{and} there is a $d>0$ such that $p(t)$
has for all $t\ge 0$ a distance $>d$ from $\partial M$.  By
construction, $\omega_{M,x}$ is invariant under not too large
deformations of $\{x_t\}$ and $M$.  For example, if the trajectory
contains a loop which comes close to $\Int M$ but does not encircle
$\partial M$, a small deformation of $\{x_t\}$ or $M$ might make the
loop intersect $\Int M$.  But, since this intersection comprises two
transitions of the trajectory through $\Int M$, one of which is
positive and one of which is negative, the value of $n(t)$ does not
change for large enough $t$.  Configurations with $p(t)$ tangential
to $\Int M$ can be evaluated as either of both limiting cases --
intersecting or not -- without effecting $\omega_{M,x}$.  Drastically
different counters can lead to different frequencies.  But each of
these is sharply defined.

\subsection{Invariance under filtering}
\label{sec:invariance}

It can be shown that for any bounded time series $\{x_t\}$ the
topological frequency is invariant under nearly arbitrary linear
filtering:

\noindent\textbf{Theorem 1.} \textit{Let $\{y_t\}$ be obtained from a
  bounded time series $\{x_t\}$ by linear, causal filtering,
\begin{align}
  \label{linear_filter}
  y_t:=\sum_{k=0}^{\infty} a_k\, x_{t-k}.
\end{align}
Assume that, for some $r>1$,
\begin{align}
  \label{positive_on_circle}
  0<\left|\sum_{k=0}^{\infty} a_k\,z^k\right|<\infty
\end{align}
for all complex $z$, $r^{-1}<|z|<r$.  (This excludes, for example,
filters which fully block some frequencies.)  Let $M$ be a counter and
$\omega_{M,x}$ be defined.  Then there is, at sufficiently high
embedding dimension, a counter $M^\prime$ such that
$\omega_{M^\prime,y}$ is defined and
$\omega_{M^\prime,y}=\omega_{M,x}$.}
  
\noindent\textbf{Proof.} This is most easily seen by the following explicit construction of an
appropriate counter $M^\prime$: Notice that the filter $\{a_k\}$ has a (not
necessarily causal) inverse $\{b_j\}$ given by
\begin{align}
  \label{inverse_series}
  \sum_{j=-\infty}^{\infty} b_j\,z^j:=
  \left(
    \sum_{k=0}^{\infty} a_k\,z^k
  \right)^{-1},
\end{align}
$r^{-1}<|z|<r$.  Let $C$ be an upper bound for $|x_t|$ and $d$ be the
(minimum) distance of the trajectory of $\{x_t\}$ from $\partial M$ in
the maximum norm.  For notational convenience define $x_t=y_t=0$ for
$t<0$.  Let
\begin{align}
  \label{define_u}
  u_t:=\sum_{j=-L}^L b_j y_{t-j},
\end{align}
where $L$ is chosen such that 
\begin{eqnarray}
\fl  |x_t-u_t|
  =
  \left|
    \left(
      \sum_{j=-\infty}^{-L-1} + \sum_{j=L+1}^{\infty} 
    \right)\sum_{k=0}^{\infty} b_j a_k x_{t-j-k}
  \right|
  \notag\\
  \lo<
  \left(
    \sum_{j=L+1}^{\infty}
    |b_j|+|b_{-j}|
  \right)
  \left(
    \sum_{k=-\infty}^{\infty} |a_k|
  \right)\,C
  \notag\\
  \label{error_estimate}
  \lo\le d/2
\end{eqnarray}
for all integer $t$.  Convergence of the left hand side of
(\ref{inverse_series}) guarantees that such an $L$ exists.
$\{u_t\}$ is an approximation of $\{x_t\}$ reconstructed from
$\{y_t\}$ using the filter~(\ref{define_u}).  Since the approximation
error of the time series is at most $d/2$, so is, in the maximum norm,
the approximation error of the trajectory.  In particular, the
topological relation between the trajectory and the counter $M$ is not
changed by going over from $\{x_t\}$ to $\{u_t\}$ (except for some
pairs of forward/backward transitions through $M$, which do not
contribute to the limit~(\ref{def_top_freq})).  Hence,
$\omega_{M,u}=\omega_{M,x}$.

Now, notice that the $N$-dimensional delay embedding of $\{u_t\}$ can
be obtained by a linear projection 
\begin{align}
  \label{define_M}
  (u_t,\ldots,u_{t-(N-1)})^T\!\!=\mathbf{P}\,(y_{t+L},\ldots,y_{t-(L+N-1)})^T
\end{align}
from the $2L+N$-dimensional delay embedding of $\{y_t\}$
\cite{broomhead92:_linear_filter_non_system,
  sauer93:_how_many__and_references}, with the matrix elements of
$\mathbf{P}$ given by (\ref{define_u}).  Furthermore, $\mathbf{P}$
maps the trajectory of $\{y_{t+L}\}$ onto the trajectory of $\{u_t\}$.
Define the oriented manifold $M^\prime$ such that
\begin{align}
  \label{define_Mp}
  \mathbf{v}\in M^\prime \Leftrightarrow \mathbf{P}\mathbf{v}\in M
  \quad (\mathbf{v}\in \mathbb{R}^{2L+N})
\end{align}
and $\partial M=\mathbf{P} \partial M^\prime$ in the obvious way.
This guarantees a finite distance of the trajectory of $\{y_t\}$ from
$\partial M^\prime$, and there is a one-to-one correspondence between
transitions of the trajectory of $\{u_t\}$ through $M$ and transitions
of the trajectory of $\{y_t\}$ through $M^\prime$.  Hence,
$\omega_{M^\prime,y}=\omega_{M,u}=\omega_{M,x}$, proving Theorem~1. $\square$

\subsection{The arbitrariness of the power spectrum, given the
  topological frequency}
\label{sec:arbitrarity}

Spectral frequency measures depend on the time series through the power
spectral density $S(\omega)$ alone.  By showing that $S(\omega)$ is
independent of the topological frequency, it becomes clear that
spectral frequency measures can generally differ arbitrarily from the
topological frequency, in contrast to what one might intuitively
assume (see, e.g., reference \cite{pikovsky97:_phase_sync}, p 226).

\noindent\textbf{Theorem 2.} \textit{Let $S_0(\omega)\ge0$ ($\omega\in [-\pi,\pi]$) be
a symmetric, continuous function, $0<\omega_0<\pi$ and $\epsilon>0$.
Then there is a time series $\{y_t\}$, an embedding dimension $N$ and
a counter $M\subset \mathbb{R}^N$ such that $\omega_{M,y}=\omega_0$
and
\begin{align}
  \label{estimate2}
  \left| S_y(\omega) - S_0(\omega) \right|<\epsilon\quad\hbox{for all $\omega\in[-\pi,\pi]$},
\end{align}
where $S_y(\omega)$ is the power spectral density of $\{y_t\}$.}

\noindent\textbf{Proof.} In order to obtain $\{y_t\}$ as described in Theorem~2,
take a time series $\{x_t\}$ which oscillates with frequency
$\omega_0$ and adjust the spectral density by filtering.
A suitable time series to start with is given by
\begin{align}
   \label{phase_x}
   x_t=2 \cos(i\, \omega_0\, t +i \,\phi_t)
   \mytext{\ with\ }
  \phi_0=0\mytext{\ and\ } \phi_{t+1}=\phi_{t}+\vartheta\, \epsilon_t,
\end{align}
where $\{\epsilon_t\}$ is an equally distributed random sequence of the
values $-1$ and $1$.  With the
counter $M$ given by (\ref{zero_crossing}), the topological
frequency $\omega_{M,x}=\omega_0$ is defined when
\begin{align}
  \label{c_condition}
  0<\vartheta<
  \left\{
    \begin{array}{ll}
      \omega_0 & \hbox{for}\quad 0<\omega_0<\pi/2,\\
      \arcsin (\sin^2\! \omega_0) & \hbox{for}\quad \pi/2\le\omega_0<\pi.\\
    \end{array}
  \right.
\end{align}
The autocorrelation function of $\{x_t\}$ is
$\left< x_t\,x_{t+\tau} \right>_t=2\, (\cos \vartheta)^\tau\cos
\omega_0 \tau$ and its spectral density
\begin{align}
  \label{Sx}
  S_x(\omega)=\frac{\sin^2\! \vartheta\,
    \left(
      1+\cos^2\! \vartheta-2 \cos \vartheta\, 
      \cos\omega\, \cos\omega_0
    \right)
    }
  {\pi
    \left|
      1-e^{i(\omega-\omega_0)} \cos \vartheta
    \right|^2
    \left|
      1-e^{i(\omega+\omega_0)} \cos \vartheta
    \right|^2
    }
\end{align}
is positive and continuous as required below.  In order to see that
there is a suitable set of filter coefficients $\{a_k\}$, notice that,
as an immediate consequence of Theorem~4.4.3 of
reference \cite{brockwell91:_time_series}, there is, for any $\epsilon>0$
and any two continuous, symmetric spectral densities $S_x(\omega)>0$
and $S_0(\omega)$ ($\omega\in[-\pi,\pi]$), a non-negative integer $p$
and a polynomial $c(z)=1+c_1 z+\ldots+c_p z^p$ such that
\begin{align}
  \label{invertable}
  c(z)\ne 0\quad\hbox{for $|z|\le 1$}
\end{align}
and, for all $\omega\in[-\pi,\pi]$,
\begin{align}
  \label{estimate1}
  \left|
    C
    \left|
      c(e^{-i \omega})
    \right|^2
    -\frac{S(\omega)}{S_x(\omega)}
  \right|<\frac{\epsilon}{\max_{\lambda}
    S_x(\lambda)},
\end{align}
where $C=(1+c_1^2+\ldots+c_p^2)^{-1} \pi^{-1} \int^\pi_0
S(\omega)/S_x(\omega) \,d\omega$.  Setting $a(z)=C^{1/2} c(z)$,
$a_0=C^{1/2}$, $a_k=C^{1/2}c_k$ ($k=1,\ldots,p$), and all other
$a_k=0$, $\{y_t\}$ given by (\ref{linear_filter}) has the spectral
density $S_y(\omega)=|a(e^{-i\omega})|^2 S_x(\omega)$ (see, e.g.,
reference \cite{brockwell91:_time_series}, Theorem~4.4.1) and
inequality~(\ref{estimate1}) implies~(\ref{estimate2}).
By~(\ref{invertable}) the filter $\{a_k\}$ satisfies the invertibility
condition~(\ref{positive_on_circle}) of Theorem~1.  Thus, an appropriate
counter $M$ can be obtained such that $\omega_{M,y}=\omega_0$ and Theorem~2
is proven.  It should be mentioned that when $S_0(\omega)$ is analytic
for real $\omega$, Theorem~2 generally holds also with the perfect
identity $S_y(\omega)=S_0(\omega)$ instead of
inequality~(\ref{estimate2}). $\square$

\subsection{An Example}
\label{sec:example}

\begin{figure}[t]
  \centering
  \epsfig{file=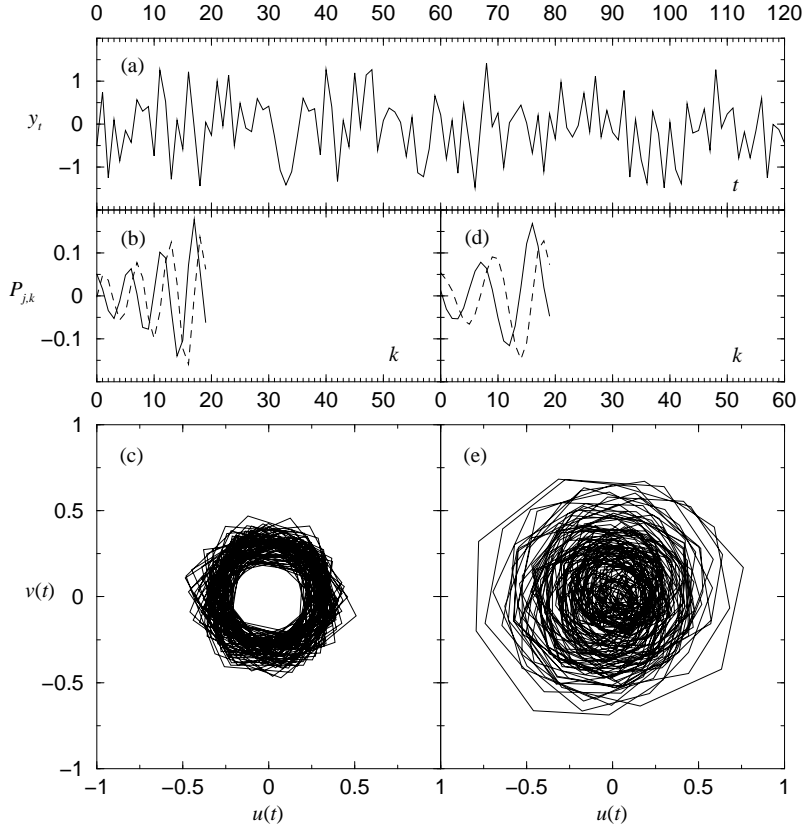,width=0.8\columnwidth}
  \caption{Demonstration for Theorem~2.  (a) A representative segment of
    the white-noise time series ${y_t}$ introduced in
    section~\ref{sec:example}, (b) the components $P_{1,k}$ (solid)
    and $P_{2,k}$ (dashed) of a projector from its 20D delay embedding
    to 2D, and (c) a segment of the projected trajectory $p(t)$ of ${y_t}$,
    with $(u(t),v(t))^T:=\mathbf{P} p(t)$.  The frequency can be read
    off.  Alternatively (d) another, inappropriate projector, and (e)
    the projected trajectory.}
  \label{fig:white}
\end{figure}
As a demonstration for Theorem~2, consider the time series $\{y_t\}$
shown in figure~\ref{fig:white}a.  By construction, it is a
realization of a white-noise process.  It was obtained by
``bleaching'' \cite{theiler93} a realization of $\{x_t\}$ given by
(\ref{phase_x}) with $\omega_0=1.1$ and $\theta=0.4$, i.e., the
realization was filtered such as to transform the known spectral
density (\ref{Sx}) into a white spectrum.  Although all spectral
information was lost, $\omega_0$ can precisely be recovered from
$\{y_t\}$.  Using an automated search algorithm (to be described
elsewhere), a projection matrix $\mathbf{P}$ (figure~\ref{fig:white}b)
is found, such that the projection of the 20D embedding of $\{y_t\}$
into 2D yields a trajectory with a nice circular structure and a
``hole'' in the center (figure \ref{fig:white}c).  The number of
oscillations and the frequency are obvious; $\omega_0$ is recovered.
Any line extending from the origin to infinity can serve as a counter
$M$ in the 2D projection.  This can be used to obtain a corresponding
counter $M'$ in 20D by a back-projection as in~(\ref{define_Mp}).

A projection with an inadequate $\mathbf{P}$ would not yield a
different frequency, but only a criss-cross kind of trajectory
(figure~\ref{fig:white}d,e), typically with an approximately Gaussian
distribution of values with a maximum density at the origin.  From
such a representation, no positive topological frequency can be
obtained.

The two-step procedure used here to obtain the counter $M'$ \emph{via}
a counter $M$ in 2D works for many experimental time series, even
though the concept of topological frequency is more general.  The
projector $\mathbf{P}$ can then be interpreted as a complex-valued
filter $\{f_k\}$ with the impulse response function
$f_k=P_{1,k}+i\,P_{2,k}$.  In section~\ref{sec:realvalued} we come
back to this point.

\section{Noisy, weakly nonlinear oscillations}
\label{sec:imperfect}

There are two assumptions upon which Theorem~1 is based -- the
boundedness of $\{x_t\}$ and the finite distance of its trajectory
from $\partial M$ -- which are not perfectly satisfied by typical
noisy processes.  Rather, the probability of reaching some point in
delay space decreases exponentially (or faster) with the distance from
some ``average'' trajectory and the inverse noise strength.  For many
processes the two assumptions and, as a consequence, the invariance of
the $\omega_{\mytext{count}}$ under filtering hold therefore only up
to an exponentially small error.  For signals generated by
\emph{noisy, weakly nonlinear oscillators}, an analytic estimate
of this error shall now be derived.

Due to the separation of time scales inherent in the weakly nonlinear
limit, it is more appropriate to work in a continuous-time
representation.  Consider the noisy, weakly nonlinear oscillator
described by a complex amplitude $A(t)$ with dynamics given by the
noisy Landau-Stuart equation
\cite{risken89:_fokker_planc_equat__chap12}
\begin{align}
  \label{Hopf_normal_form}
  \dot A = (\epsilon +i \omega_0) A - (1+i g_i)|A|^2 A + \eta(t),
\end{align}
where $\epsilon$, $\omega_0$, and $g_i$ are real and $\eta(t)$
denotes complex, white noise with correlations
\begin{align}
  \label{white_noise}
    \left< \eta(t)\eta(t') \right>=0,\quad
    \left< \eta(t)\eta(t^\prime)^*
    \right>=4 \delta(t-t^\prime)
\end{align}
[$^*$ $\equiv$ complex conjugation, $\left<\cdot \right>$ $\equiv$
expectation value].  In a certain sense, this
system universally describes noisy oscillations in the vicinity of a
Hopf bifurcation \cite{arnold98:_random_dynam_system}.

\subsection{Definitions of frequency}
\label{sec:other-definitions}

\begin{figure}
  \begin{center}
    \epsfig{file=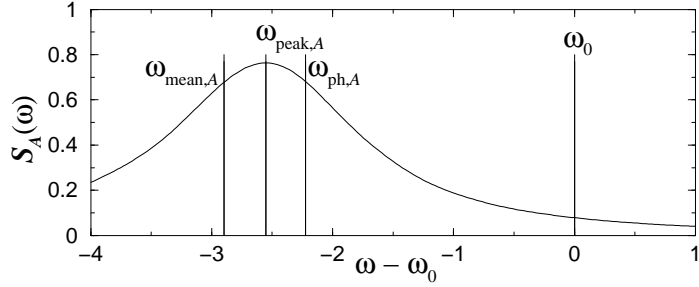,width=0.7\columnwidth}
  \end{center}
  \caption{The power spectrum $S_A(\omega)$ of $A(t)$ given by
    equations\ (\ref{Hopf_normal_form},\ref{white_noise}) with $\epsilon=2$
    and $g_i=1$, obtained from a numerical simulation; and
    $\omega_{\mytext{peak},A}$ (analytic results in
    reference \cite{seybold74:_theor_detun_singl_mode_laser_near_thres}),
    compared to the mean frequency $\omega_{\mytext{mean},A} = \omega_0
    - g_i\left<|A|^4\!\right>/\left<|A|^2\!\right>$ [$\left< |A|^{2n}
      \!\right>=2^n\,\mathcal{N}^{-1}\,d^n\!\mathcal{N}/d\epsilon^n$,
    see (\ref{normalization_constant})] and the phase frequency
    $\omega_{\mytext{ph},A}=\omega_0- g_i \left< |A|^2 \!\right>$,
    defined by equations\ (\ref{define_f_phase},\ref{define_f_mean}), and
    the linear frequency $\omega_0$.}
  \label{fig:peak_vs_phase_f}
\end{figure}

For the reasons explained above, the topological frequency cannot be
defined rigorously for noisy, weakly nonlinear oscillators.  The
customary frequency measures, such as the linear frequency $\omega_0$,
the spectral peak frequency $\omega_{\mytext{peak}}$, the average
frequency or \emph{phase frequency}
\begin{align}
  \label{define_f_phase}
  \omega_{\mytext{ph},A}:=
  \left<
    \omega_i
  \right>,\quad \omega_i:=\mathrm{Im}\left\{\smash[t]{\dot A}/A\right\},
\end{align}
and
the \emph{mean frequency}
\begin{align}
  \label{define_f_mean}
  \omega_{\mytext{mean},A}:=
  \frac{  
    \left<
      \omega_i |A|^2
    \right>
    }{\left<
      |A|^2
    \right>
    }
  = \frac{
    \mathrm{Im}
      \left<
        \smash{\dot A} A^*
      \right>
      }{
    \left<
      |A|^2
    \right>}
  =\frac{\int \omega S_{
      A}(\omega)d\omega}{\int S_{A}(\omega)d\omega}
\end{align}
will generally (i.e., with $g_i\ne 0$) all yield different values; see
figure~\ref{fig:peak_vs_phase_f}.
[Definitions~(\ref{define_f_phase},\ref{define_f_mean}) are sometimes
restricted to ``analytic signals'' ($S_A(\omega)=0$ for $\omega<0$)
derived from the corresponding real-valued signals
$\mathrm{Re}\{A(t)\}$.  See reference
\cite{boashash92:_estim_inter_instan_frequen} for the history.]

The phase frequency measures the average number of circulations around
the point $A=0$ in phase space per unit time (decompose
$A(t)=a(t)e^{i\phi(t)}$ to see this).  It is a period-counting
frequency and the quantity which comes conceptually closest to the
topological frequency.  However, the choice of the point $A=0$ can
here be justified only by symmetry and dynamics [the invariant density
pertaining to equation~(\ref{Hopf_normal_form}) has an extremum at $A=0$],
and not by invariance under perturbations.  $\omega_{\mytext{peak}}$ and
$\omega_{\mytext{mean}}$ are both spectral frequency measures, and the
influence of filtering is obvious.  But how does filtering
affect $\omega_{\mytext{ph},A}$?  The following considerations lead to
a surprisingly accurate result.

\subsection{The effect of filtering on the phase frequency}
\label{sec:filtering}

The dynamics of $A$ on short time scales $\delta t$ is dominated by
the driving noise, and the change in $A$ is of the order $|\delta
A|=\mathcal{O}(4 \delta t)^{1/2}$.  A band-pass filter of spectral width
$\Delta \omega$ which truncates the tails of the peak corresponding to
$A$ in the power spectrum suppresses this diffusive motion on time
scales $\Delta \omega^{-1}$, while on longer time scales dynamics
change only little.  The corresponding deformation of the path of $A$
in the complex plane can alter the number of circulations of the
origin whenever $A$ approaches the origin to less then
$\approx(4/\Delta\omega)^{1/2}$.  At these times $|A|$ is small and, for
not too narrow filters, the dynamics of $A$ in its linear range.
Thus, the effect of broad-band filtering can be estimated by a linear
theory.  Consider, for a moment,  the linearized version of
equation~(\ref{Hopf_normal_form}),
\begin{align}
  \label{linear_oscillator}
  \dot { A} = (\epsilon +i \omega_0)  A + \eta(t),
\end{align}
with $\eta(t)$ as above, and assume $\epsilon<0$.
Clearly, $\omega_{\mytext{ph}, A}=\omega_0$.  For the phase frequency
of a complex, Gaussian, linear process $B(t)$ in general, a simple
calculation shows
$\omega_{\mytext{ph},B}=\omega_{\mytext{mean},B}$.
This
can be used to calculate the phase frequencies of $ A$ after
filtering.  Let, for example, $B$ be obtained from $A$ through the
primitive band-pass filter
\begin{align}
  \label{filtered_linear}
  \dot { B}=(-\epsilon_1+i \omega_1) B + A,
\end{align}
which is centered at $\omega_1$ with width $\epsilon_1>0$.  Using
$\omega_{\mytext{ph},B}=\omega_{\mytext{mean},B}$ and elementary
filter theory \cite{priestley81:_spect} one obtains
\begin{align}
  \label{linear_f_phase1}
  \omega_{\mytext{ph}, B}=\frac{\epsilon_1\, \omega_0-\epsilon
    \,\omega_1}{\epsilon_1-\epsilon}.
\end{align}

\begin{table}[b]
  \caption{\label{tab:effect_of_filtering}
    The shift $\delta \omega_{\rm{num}}$ in the phase
    frequency  $\omega_{\mytext{ph},A}$ of $A(t)$, obtained
    from simulations of
    equations~(\ref{Hopf_normal_form},\ref{white_noise}) with $g_i=1$, 
    after filtering
    as in (\ref{filtered_linear}),  and a 
    comparison with the theoretical estimate (\ref{delta_omega}).  
    The data verify 
    $\delta \omega\sim\epsilon_1^{-1}$,
    $\sim(\omega_1-\omega_0)$, 
    and $\sim\mathcal{N}^{-1}$ 
    in this order.}
  \begin{indented}
    \item[]
    \begin{tabular}{ccccccc}
      \br
      $\epsilon$ & $(\omega_{\mytext{ph},A}-\omega_0)$ & $(\omega_1-\omega_0)$ 
      & $\epsilon_1$ &
      $\delta \omega_{\rm{theo}}$ & $\delta \omega_{\rm{num}}$ \\
      \mr
      2 & $-2.2253$ & $-2.5$ & 48 & $-0.0117$ & $-0.0109(16)$ \\
      2 & $-2.2253$ & $-2.5$ & 24 & $-0.0235$ & $-0.0228(16)$ \\  
      2 & $-2.2253$ & $-2.5$ & 12 & $-0.0469$ & $-0.0437(15)$ \\
      2 & $-2.2253$ & $\phantom{-}0.0$ & 24  & $\phantom{-}0.0000$ & $ \phantom{-}0.0003(17)$ \\ 
      2 & $-2.2253$ & $\phantom{-}2.5$ & 24 & $\phantom{-}0.0235$ & $\phantom{-}0.0226(16)$ \\
      0 & $-1.1284$ & $-2.5$ & 24 & $-0.1175$ & $-0.1036(47)$ \\
      3 & $-3.0605$ & $-2.5$ & 24 & $-0.0063$ & $-0.0074(13)$ \\
      4 & $-4.0104$ & $-2.5$ & 24 & $-0.0011$ & $-0.0012(11)$ \\
      \br
    \end{tabular}
\end{indented}
\end{table}
By the argument given above, the shift in phase frequency
$\delta\omega:=\omega_{\mytext{ph}, B} -
\omega_{\mytext{ph}, A} =
(\omega_0-\omega_1)\,(\epsilon/\epsilon_1)+\mathcal{O}(\epsilon_1^{-2})$
is due to the times where $| A|^2 \lesssim 4/\epsilon_1$.
Since $ A$ has a complex normal distribution with variance
$\big<| A|^2\big>= -1/\epsilon$, this happens about
\begin{align}
  \label{linear_center_rate}
  p\left[| A|^2<\frac{4}{\epsilon_1}\right]=1-\exp
  \left(
    \frac{2 \epsilon}{\epsilon_1}
  \right)=-\frac{2 \epsilon}{\epsilon_1}+\mathcal{O}(\epsilon_1^{-2})
\end{align}
of all times.  Thus, during these times, the shift in phase frequency
is $\delta \omega/p[| A|^2<{4}/{\epsilon_1}]=(\omega_1-\omega_0)/2
+\mathcal{O}(\epsilon_1^{-2})$.  Extrapolation to $\epsilon>0$ and the
weakly nonlinear case yields
\begin{align}
  \label{nl_frequency_shift}
  \delta \omega=p\left[| A|^2<\frac{4}{\epsilon_1}\right]
  \frac{\omega_1-\omega_0}{2}+\mathcal{O}(\epsilon_1^{-2}),
\end{align}
now with 
\begin{align}
  \label{nl_p}
  p[|A|^2<I_0]=\frac{1}{\mathcal{N}}\int_0^{I_0}\exp\left(\displaystyle
    \frac{\epsilon I}{2}-\frac{I^2}{4}\right)\,dI=\frac{I_0}{\mathcal{N}}+\mathcal{O}(I_0^2),
\end{align}
where
\begin{align}
  \label{normalization_constant}
  \mathcal{N}=\pi^{1/2}\exp(\epsilon^2/4) [1+\erf(\epsilon/2)].
\end{align}
Equations~(\ref{nl_frequency_shift}-\ref{normalization_constant})
predict the shift 
\begin{align}
  \label{delta_omega}
  \delta \omega=2\epsilon_1^{-1} \mathcal{N}^{-1}(\omega_1-\omega_0) +
  \mathcal{O}(\epsilon_1^{-2})
\end{align}
in the phase frequency of $A(t)$ given by
(\ref{Hopf_normal_form},\ref{white_noise}) after passing through the
filter~(\ref{filtered_linear}).  A numerical test verifying this
result is shown in table~\ref{tab:effect_of_filtering}; notice in
particular the fast decay of $\delta \omega$ as $\epsilon$ increases
[$\delta \omega\sim\exp(-4/\epsilon^2)$] and the conditions of
Theorem~1 are better satisfied.  The high accuracy of the result might
be understood by observing that the crude upper bound
$|A|^2<{4}/{\epsilon_1}$ enters the derivation of (\ref{delta_omega})
two times, its numerical value canceling out.

\subsection{The frequency of real-valued, weakly nonlinear signals}
\label{sec:realvalued}

Experimental signals are real-valued.  Assume that, instead of $A(t)$,
only a real-valued signal $x(t)=\mathrm{Re}\{A(t)+(\hbox{higher
  harmonics})\}+(\hbox{perturbations})$ is given.  The natural way to
estimate the phase frequency of $A(t)$ then is to construct an
approximation $\hat A(t)=(f*x)(t)$ of $A(t)$ by a convolution of
$x(t)$ with a complex-valued filter $f(t)$, and to estimate the phase
frequency as $\hat \omega_{\mytext{ph},A}=\omega_{\mytext{ph},\hat
  A}$.  The filter $f(t)$ describes the combined effect of 2D delay
embedding or analytic-signal construction, and filtering to eliminate
higher harmonics, offsets, aliasing, and external perturbations.  The
result above shows that generally, for $\hat \omega_{\mytext{ph},A}$
to be unbiased, the total effect of all these transformations should
be a complex, symmetric band-pass centered on the linear frequency
$\omega_0$ ($\ne \omega_{\mytext{ph}},\,\omega_{\mytext{peak}}$!).
Otherwise there is a bias which decays as $\exp(-4/\epsilon^2)$ for
large $\epsilon$.  To the extent that the bias vanishes, the
probability of finding values of $\hat A(t)\approx A(t)\approx 0$
vanishes, too.  Then, a counter $M'$ for $x(t)$ can be obtained along
the lines of section~\ref{sec:example} using the filter $f(t)$ --
approximated by a time-discrete filter $f_k$ -- for the projection to
2D.  Obviously, the corresponding topological frequency
$\omega_{M',x}$ equals $\hat \omega_{\mytext{ph},A}$.

\section{Conclusion}
\label{sec:conclusion}

When the spectral density is of genuine interest, forget period
counting.  But there are many real-world applications (e.g.,~ in
astronomy \cite{mendez98:_differ_frequen_kiloh_qpos}, earth science
\cite{godano99:_sourc_strom_vulcan_islan_isole_eolie_italy},
biomedicine
\cite{timmer96:_quant_tremor,korhonen02:_estim_cardiovasc}, or
engineering \cite{slavic02:_measur_turbo_wheel}) where neither the
characteristics of the signal pathway nor a detailed model of the
oscillator are known, and yet a robust measure of the frequency or, at
least, some robust characterization of the oscillator is sought.
Then, by Theorem~2, spectral methods miss valuable information.  In
view of Theorem~1 and equation~(\ref{delta_omega}), concepts such as
topological frequency or its little brother, phase frequency, are more
appropriate.  The fractal dimension of the reconstructed attractors,
an alternative characterization, is typically robust with respect to
finite-impulse-response filtering only
\cite{broomhead92:_linear_filter_non_system,
  sauer93:_how_many__and_references}, i.e., only if there is a $q$
such that $a_k=0$ for $k>q$ in (\ref{linear_filter}).  

For a practical application of topological frequency, a systematic
method to find appropriate counters in the typically high-dimensional
delay spaces is desirable.  Some progress in this direction has be
made and will be reported elsewhere.

\ack{Work supported by the German Bundesministerium f{\"u}r
Bildung und Forschung (BMBF), grant 13N7955.}

\vspace{\fill}


\end{document}